\documentclass[12pt,preprint]{elsarticle}
\usepackage{amssymb,epsfig,graphicx,multirow,bm,dcolumn,rotating}
\journal{JALCOM}
\begin{document}
\begin{frontmatter}

\title{$d$ band filling and magnetic phase separation in transition metal-doped Mn$_3$SnC}
\author[gu]{V. N. Gaonkar}
\author[tifr]{E. T. Dias}
\author[tifr]{A. K. Nigam}
\author[gu]{K. R. Priolkar \corref{krp}}\ead{krp@unigoa.ac.in}
\cortext[krp]{Corresponding author}
\address[gu]{Department of Physics, Goa University, Taleigao Plateau, Goa 403206 India}
\address[tifr]{Tata Institute of Fundamental Research, Dr. Homi Bhabha Road, Colaba, Mumbai 400005 India}

\begin{abstract}
The structural and magnetic properties of transition metal-doped Mn$_3$SnC are studied with an aim to understand the effect of transition metal atom on magnetostructural properties of the antiperovskite compound. The doped  Mn$_{2.8}$T$_{0.2}$SnC (T = Cr, Fe, Co, Ni and Cu) compounds show a distinctly different magnetic behavior which can be related to electronic filling of the $d$ band of the transition metal atom. While Cr and Fe doped Mn$_3$SnC show properties similar to that of Mn$_3$SnC, the Co, Ni and Cu doped compounds exhibit nucleation of secondary phases which are devoid of carbon and having Heusler and DO19 type hexagonal structure.  A strong magnetic interaction is observed between the impurity phases and the major antiperovskite phase leading to a sharp decrease in magnetostructural transition temperature of the antiperovskite phase and a cluster glassy ground state.
\end{abstract}
\date{\today}

%\maketitle
\end{frontmatter}

\section{Introduction}

The antiperovskite compounds with general stoichiometry Mn$_3$AX, where A site is mainly occupied by the main group and transitional elements and X site by C, N or B etc. exhibit a numerous interesting physical properties like near-zero temperature coefficient of resistance \cite{takenaka201198e,Lei201199n}, giant magnetoresistance (GMR) \cite{Kamishima200063g,Li200572m,Zhang2014115m}, giant negative thermal expansion \cite{Hamada2011109g,Takenaka200587g}, large magnetocaloric effect (MCE) \cite{Tohei200394n,yu200393L,Aczel201490}, piezo-magnetic effect \cite{lukashev200878t}, etc. The presence of such interesting properties and a chemically flexible structure that accommodates different elements into its matrix have made these antiperovskite compounds attract attention.

In the past decade the antiperovskites, Mn$_3$GaC and Mn$_3$SnC have been studied extensively. Among the two, Mn$_3$SnC undergoes a first-order magnetostructural transition from high temperature paramagnetic (PM) state to the low temperature  magnetically ordered state with complex magnetic order near room temperature (T $\sim$ 280 K) \cite{dias201548ef} accompanied by a large entropy change -$\Delta$S$_{M}$ of the order of 8 J/kg K in an applied magnetic field of 20 KOe \cite{wang200985L}. The presence of such large magnetocaloric effect in vicinity of room temperature makes Mn$_3$SnC a suitable material for the magnetic refrigeration.

Site-specific doping is considered as an effective strategy to alter and improve physical properties of a compound. In antiperovskites, for instance, N doping for C in Mn$_3$GaC, though does not increase the first-order magnetic transformation temperature, reduces the transformational hysteresis considerably \cite{ccakir2013344adi}. A similar reduction of transformational hysteresis is also seen in Sn doped Mn$_3$GaC \cite{dias2015117ef}. Moreover, such studies shed light on the nature and the cause of magnetic interactions in Mn-based antiperovskites. In Mn$_3$Ga$_{1-x}$Sn$_x$C, the local structural distortions of the Mn$_6$C octahedra within the cubic cage formed by the Ga/Sn atoms are shown to be responsible for antiferromagnetic order. Further, the strain introduced by the A site atom, (Ga or Sn), plays a vital role in modulating the magnetic interactions as well as nature of the magnetocaloric effect \cite{dias2018124ph}. The local structural distortions are so intimately dependent on the A site atom that despite being single phase cubic in structure, Mn$_3$Ga$_{0.5}$Sn$_{0.5}$C displays two different types of magnetic orders due to its local magnetic phase separation into Ga rich and Sn rich regions \cite{dias201795ph}.

The majority of the work so far, is primarily focused on substitution of A or X site elements in Mn$_3$AX antiperovskite compounds \cite{dias20141co,kanomata199067p,kamishima1997237m,wang201097m,sun201093Ne}. Very few reports present studies on substitution at the Mn site. Such substitution will directly affect the magnetic interactions and could have dramatic effect on magnetostructural transition and the associated magnetocaloric effect \cite{harada199923Ma,kanomata19937Ma,wang2010405St}. For instance, Doping small quantities of Fe, Co and Ni at the Mn site in Mn$_3$GaC has been found to improve the ferromagnetic transition temperature $T_{C}$, while Cr substitution at the Mn site results in decrease in $T_{C}$ \cite{harada1995140Ma}.

Mn$_3$SnC has a peculiar magnetic order. Even though the crystal structure is cubic, magnetically one of the three equivalent Mn atoms carries ferromagnetic moment while the other two align antiferromagnetically with propagation vector, $k = [{1 \over 2}, {1 \over2}, 0$] \cite{dias201548ef}. Time-dependent magnetization measurements have revealed that at the magnetostructural transformation, initially all three Mn atoms order ferromagnetically and with time two of them flip and align antiferromagnetically \cite{ccakir201796Dy}. The relaxation time of these spin flipping competes with the lattice relaxation time and can be effectively modulated by introduction of C or Sn vacancies \cite{gaonkar2019471Ro}. Substitution of Mn atoms by other transition metal atoms can directly affect the nature of magnetic interactions as well as the flipping of magnetic spins. In fact at low doping levels ($x \le 0.2$), replacing Mn by Fe in Mn$_{3-x}$Fe$_x$SnC has been shown to an effective strategy to manipulate the properties of the antiperovskite compound \cite{wang2010322m}. However, higher Fe content $0.5 \le x \le 1.3$ leads to a strong competition between FM and AFM components and a field-induced ferromagnetism \cite{nie2011151m}. It would, therefore, be interesting to systematically probe the effect of transition metal substitution for Mn on the magnetic and structural properties of Mn$_3$SnC. For the present study, we have chosen transition metal atoms from Cr through Cu as dopants for Mn in Mn$_{3-x}$T$_x$SnC at a fixed doping level of $x = 0.2$. Our results show two distinct effects as a function of transition element doping. The compounds doped with transition metal atoms with a lower $d$-band filling like Cr and Fe form structurally single-phase compounds. On the other hand, compounds doped with Co, Ni and Cu (higher $d$-band filling) exhibit structural and magnetic phase separation. The two magnetic phases interact to completely modify the properties of the major antiperovskite phase and result in a cluster glassy ground state.

\section{Experimental}
A series of five polycrystalline Mn$_{2.8} $T$_{0.2}$SnC (T = Cr, Fe, Co, Ni and Cu) metallic antiperovskites samples were prepared along with Mn$_{3}$SnC by employing the solid-state reaction method. To begin with the weights of constituent elements, Mn:T:Sn:C, in the ratio 2.8:0.2:1:1 were calculated and accordingly weighed. A 15\% excess carbon by weight was added to ensure C stoichiometry in the final compounds \cite{dias201548ef}.  The elements were then powdered and mixed thoroughly using a mortar and pestle to form a homogeneous mixture, pressed in to pellets of diameter 10 mm and sealed individually in an evacuated quartz tube. The sealed tubes were subjected to sintering for 48hrs at 1073K and then for 120hrs at 1150K. Later upon furnace cooling to room temperature, the samples were powdered and annealed again at the same temperature to ensure homogeneity.  The phase identification in the resulting compounds was done by recording x-ray diffraction (XRD) patterns on each of these samples and performing LeBail refinement. SEM-EDX measurements were also performed to verify composition and surface morphology. Magnetization as a function of temperature was recorded on all prepared samples using  MPMS-SQUID magnetometer. Magnetization was recorded in a field of 0.01 T while increasing the temperature from 5K to which they were cooled in zero applied field (ZFC) as well as during subsequent cooling (FCC) and warming (FCW) cycles. Isothermal magnetization was recorded in a magnetic field interval of $\pm$ 7 T at 300K and 5K. For these measurements the samples were cooled to the desired temperature in zero applied field. The relaxation dynamics in all the samples were studied from time-dependent magnetization measurement. For these measurements, the sample was cooled in zero applied field to 5K. A waiting time $t_w$ was used to equilibrate the spin system before applying a magnetic field of 0.01 T and recording magnetization as a function of observation time $t$. The ac magnetic susceptibility measurements were performed using a Physical Properties Measurement System (Quantum Design). Measurements in the 2 -- 300 K temperature range were carried out at various excitation frequencies (33 $\le$ f $\le$ 10000 Hz) by applying an ac amplitude $H_{ac}$ = 10 Oe after cooling the sample in zero field.

\section{Result and Discussion}
\begin{figure}
\begin{center}
\includegraphics[width=\columnwidth]{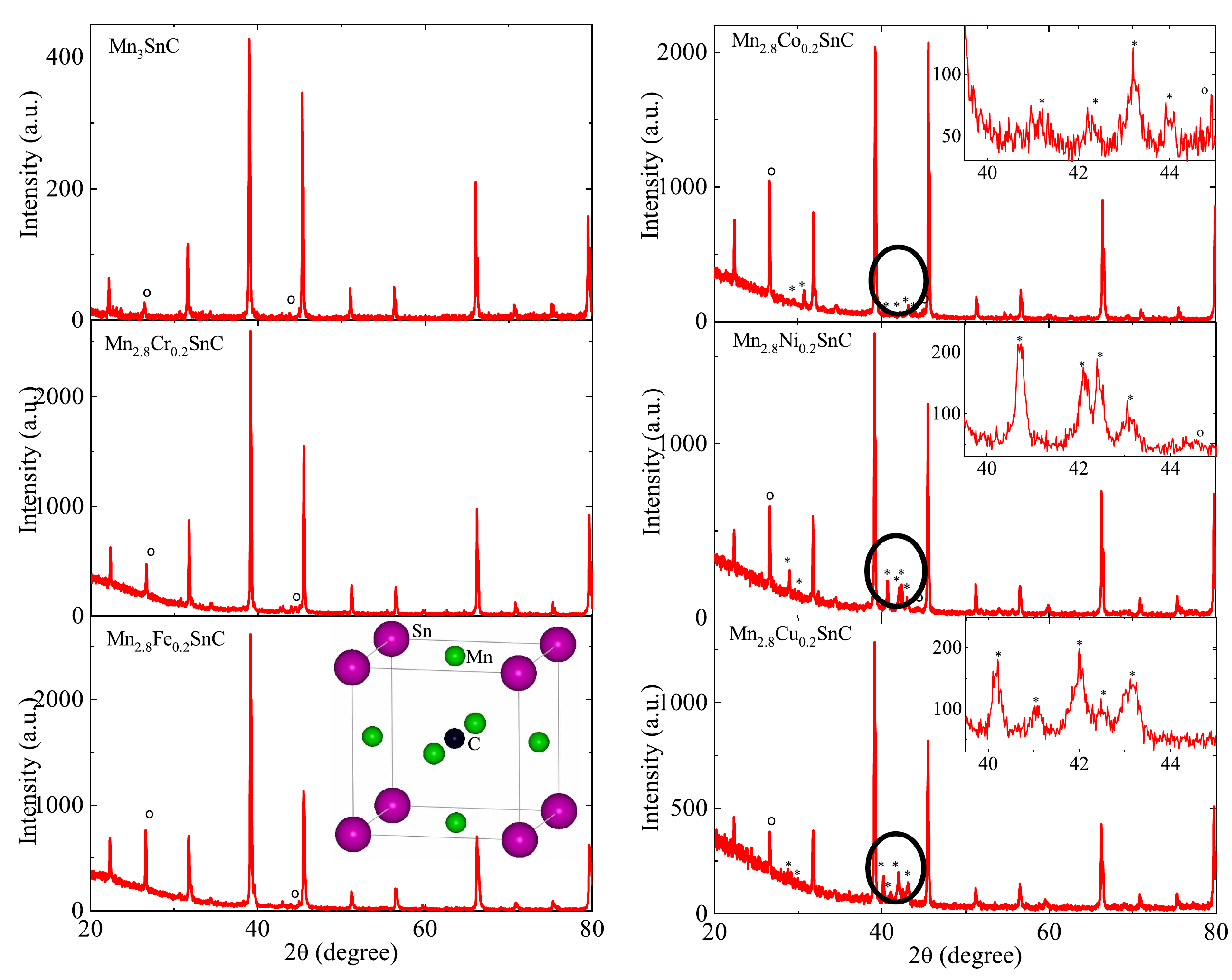}
\caption{(a--f) XRD patterns of Mn$_{3}$SnC,  Mn$_{2.8}$Cr$_{0.2}$SnC, Mn$_{2.8}$Fe$_{0.2}$SnC, Mn$_{2.8}$Co$_{0.2}$SnC, Mn$_{2.8}$Ni$_{0.2}$SnC and Mn$_{2.8}$Cu$_{0.2}$SnC polycrystalline samples respectively. Insets in (d -- f) show the expanded view of circled region. The symbols 'o' and '*' indicates the reflection corresponding to the impurity phases. An antiperovskite unit cell is shown as inset in (c). The doped transition metal atoms are expected to replace Mn atoms at the face centre positions.}
\label{fig:1TMXRD}
\end{center}
\end{figure}

The XRD patterns of Mn$_{2.8}$T$_{0.2}$SnC (T = Cr, Fe, Co, Ni and Cu) compounds along with stoichiometric Mn$_{3}$SnC compound are depicted in figure \ref{fig:1TMXRD}. Among the five, Cr and Fe doped compounds show almost a single-phase XRD pattern similar to that of Mn$_{3}$SnC with only a small percentage ($\sim 2\%$) of C impurity. On the other hand the compounds doped with Co, Ni and Cu present multiphasic compounds with more than one impurity phases along with the main antiperovskite phase and the C phase. At the current doping level it is difficult to identify the impurity phases unambiguously. Hence additional two Ni-doped compounds, Mn$_{2.5}$Ni$_{0.5}$SnC and Mn$_3$NiSnC were prepared and, their diffraction patterns were analyzed.

\begin{figure}
\begin{center}
\includegraphics[width=\columnwidth]{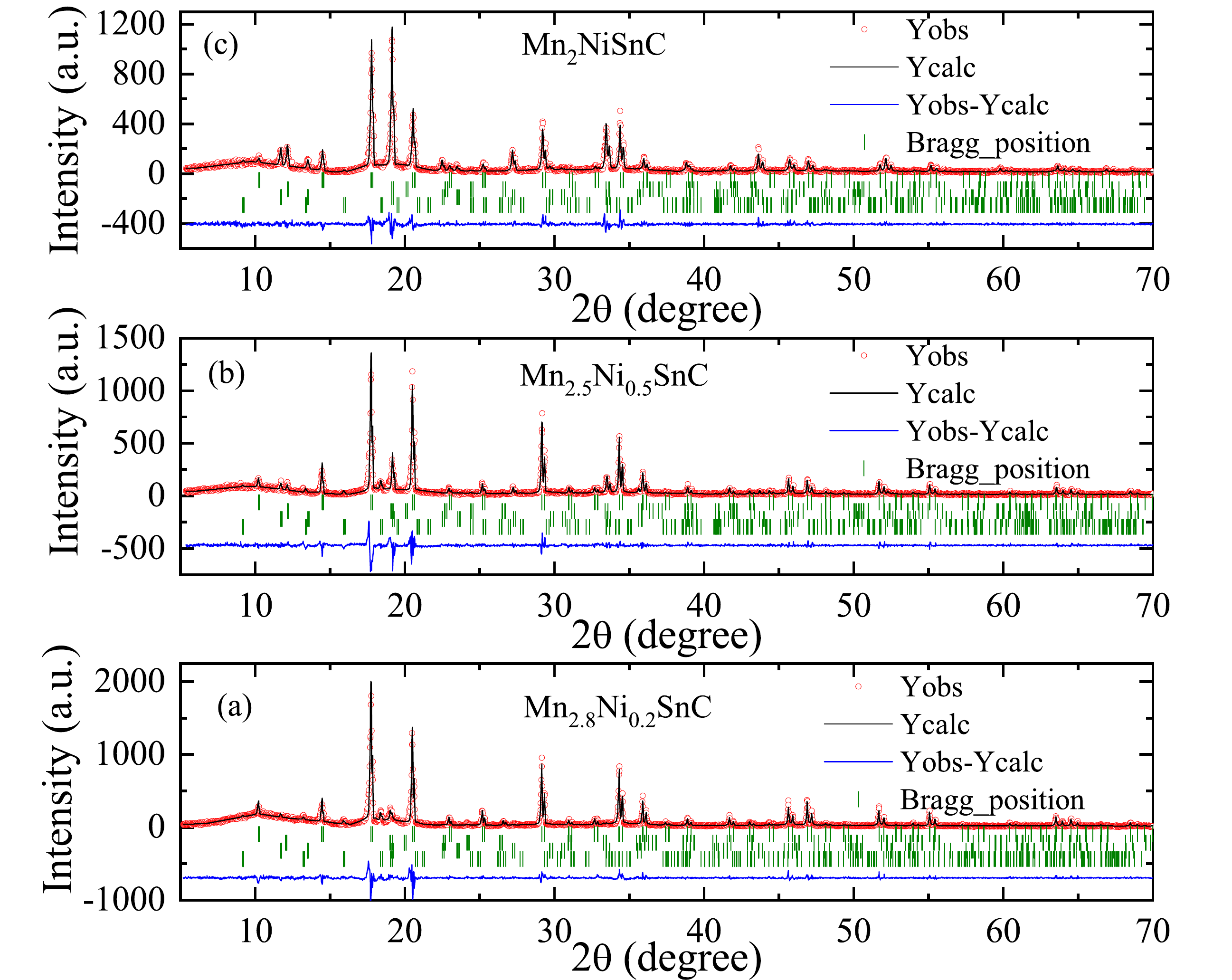}
\caption{(a)--(c) Le-Bail refined XRD patterns of Mn$_{3-y}$Ni$_y$SnC ($y$ = 0.2, 0.5 and 1.0)  polycrystalline samples recorded using Mo target ($\lambda$ = 0.709 \AA). Phases I to IV respectively are cubic antiperovskite phase, graphite phase, Heusler phase, and DO19 phase.}
\label{fig:2XRD-Mn2NiSnC}
\end{center}
\end{figure}

Le Bail refined XRD patterns of the three Ni-doped compounds are presented in figure \ref{fig:2XRD-Mn2NiSnC}. A total of four phases could be identified, the antiperovskite phase, cubic Heusler phase, DO19 type hexagonal phase and the hexagonal graphite phase. As can be seen from figure \ref{fig:2XRD-Mn2NiSnC}, these four phases describe the observed diffraction patterns quite well. Although it is difficult to determine the chemical composition of the observed phases using x-rays as the constituent elements include Mn (Z = 25) and Ni (Z = 28), a simple exercise of dividing the constituent elements into the observed four phases indicates the Heusler and DO19 phases to most likely be Mn$_2$NiSn and Ni$_3$Sn respectively. This simple exercise implies that the doped transition metal atom does not substitute Mn in the antiperovskite phase. Therefore, for instance, the compound like Mn$_2$NiSnC could segregate into Mn$_3$SnC, Mn$_2$NiSn, Ni$_3$Sn and graphitic carbon. Such a distribution of impurity phases would imply a decrease in the phase fraction of the antiperovskite phase concomitant with an increase in the phase fraction of carbon phase with the increase in Ni doping content. Indeed one can observe an increase in intensity of (002) Bragg reflection corresponding to hexagonal graphite phase and a decrease in the intensity of Bragg reflections of the antiperovskite phase as the Ni content is increased from 0.2 to 1.0. The increase of about 33\% in the carbon phase and the corresponding decrease of $\sim$65\% in the antiperovskite phase agrees well with the phase distribution suggested above. A similar distribution of four phases could be possible in the Co and Cu doped Mn$_3$SnC. Cu$_3$Sn, Co$_3$Sn and Mn$_2$CoSn are well known compounds \cite{Sang2009469or,wijn1991alloys,lakshmi200225hy} while Mn$_2$CuSn have been recently predicted to crystallize in tetragonal structure \cite{faleev20171He}. The lattice parameters of antiperovskite phase, obtained via Le Bail refinement, in all transition metal-doped antiperovskites are compared with that of Mn$_3$SnC in Table \ref{tab1}. The values of lattice parameter do not show a systematic trend that could be related to size of the doped transition metal atom. In fact, it is observed that lattice parameters, initially decrease, from T = Cr to T = Co and then increase again. The initial decrease in lattice parameter from Cr to Co augers well with variation of atomic radius of these atoms. The increase in lattice parameter in Ni doped compound however, is not expected. This increase in lattice parameter along with observation of impurity phases lends support to the argument that transition metal atoms like Co, Ni and Cu do not replace Mn in the antiperovskite phase, instead form Heusler and DO19 type impurity phases.

%\begin{sidewaystable}
\begin{table*}
\centering
\caption{\label{tab1} Lattice constant of the antiperovskite phase, transformation temperature, magnetic moment and EDX compositions observed in Mn$_{3-x}$T$_x$SnC (T = Cr, Fe, Co, Ni and Cu) and $x$ = 0.2}
%\begin{tabular}{cccccc}
\begin{tabular*}{\textwidth}{@{}cccccc}
\hline
Composition & Lattice  & Transformation  & Magnetization & \multicolumn{2}{c}{EDX Composition}\\
& Constant (\AA) & temperature ($T_{ms}$) (K) & at 7 T (emu/g) & Major Phase	& Minor Phase\\
\hline
Mn$_3$SnC & 3.9950(1) &	281	& 22.22873 & Mn - 74.5 & --\\
& & & & Sn - 25.5 &	--\\
\hline
Mn$_{2.8}$Cr$_{0.2}$SnC & 3.9893(1) & 280 & 20.8812 & Mn - 69.5 & -- \\
&&&& Sn - 23.5 & \\
&&&& Cr - 7.0 & \\
\hline	
Mn$_{2.8}$Fe$_{0.2}$SnC &	3.9855(1) &	279	& 21.66341 &	Mn - 74.0 & -- \\
&&&& Sn - 19.1 &\\
&&&& Fe - 6.9 &\\
\hline
Mn$_{2.8}$Co$_{0.2}$SnC &	3.9763(1) & 209 & 17.58977 &	Mn - 69.3 & Mn - 59.3\\
&&&& Sn - 25.7 & Sn - 25.7\\
&&&& Co - 5.0 & Co - 15.0\\
\hline
Mn$_{2.8}$Ni$_{0.2}$SnC &	3.9814(1) & 234 &	26.28319 &	Mn - 80.9 & Mn - 42.0\\
&&&& Sn - 18.6 & Sn - 29.9\\
&&&& Ni - 0.5 & Ni - 28.0 \\
\hline
Mn$_{2.8}$Cu$_{0.2}$SnC &	3.9854(1) &	245 & 22.76375 & Mn - 81.1 & Mn - 46.0\\
&&&& Sn - 17.3 & Sn - 30.9\\
&&&& Cu - 1.6 & Cu - 23.1\\
\hline
\end{tabular*}
%\end{tabular}
%\end{sidewaystable}
\end{table*}

\begin{figure}
\begin{center}
\includegraphics[width=\columnwidth]{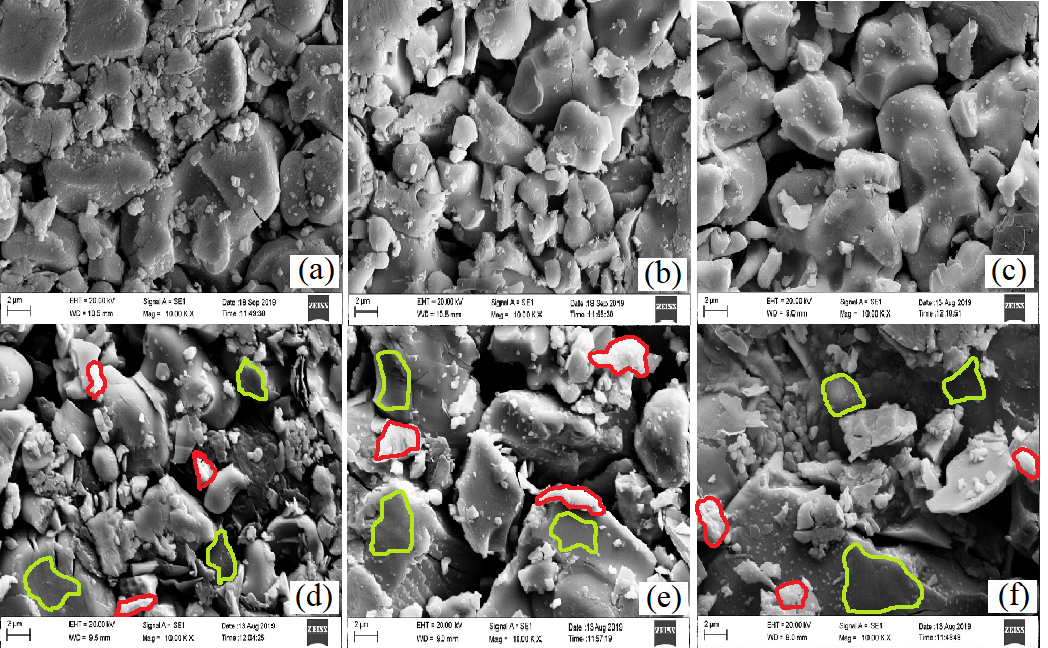}
\caption{SEM micrographs of (a) Mn$_3$SnC, (b) Mn$_{2.8}$Cr$_{0.2}$SnC, (c) Mn$_{2.8}$Fe$_{0.2}$SnC, (d) Mn$_{2.8}$Co$_{0.2}$SnC, (e) Mn$_{2.8}$Ni$_{0.2}$SnC and (f) Mn$_{2.8}$Cu$_{0.2}$SnC}
\label{fig:3SEM}
\end{center}
\end{figure}

To further confirm the presence of impurity phases and their morphological nature, SEM-EDX spectroscopy was performed on all compositions. The SEM micrographs obtained for Mn$_3$SnC and Mn$_{2.8}$T$_{0.2}$SnC (X = Cr, Fe, Co, Ni, and Cu) compounds are presented in figure \ref{fig:3SEM}. The surface morphology of Mn$_3$SnC, Mn$_{2.8}$Cr$_{0.2}$SnC and Mn$_{2.8}$Fe$_{0.2}$SnC appear quite similar (see figure \ref{fig:3SEM}(a)--(c)), while additional morphological phase is seen in Co, Ni and Cu doped compounds (figure \ref{fig:3SEM}(d)--(f)). This additional phase grows with an increase in transition element content (not shown). EDS analysis for chemical composition of these two phases indicates that the minority phase is rich in the doped transition element while the majority phase contained negligible amounts of doped transition metal. The average compositions determined for the two phases typically were Mn$_{74}$X$_2$Sn$_{24}$ and Mn$_{50}$X$_{25}$Sn$_{25}$ (X = Co, Ni or Cu) for the majority and minority phases respectively. The individual compositions for each of the samples are given in Table \ref{tab1}. The 3:1 ratio of Mn:Sn in the majority phase indicates it to be the antiperovskite phase while 2:1:1 ratio of Mn:Ni:Sn in the minority phase agreed well with the Le Bail identification of inverse Heusler phase. We could not unequivocally identify the presence of DO19 phase. The relatively small percentage of the two phases makes it difficult to differentiate between Heusler and DO19 phases but, the non-stoichiometric ratio of Mn:X in the average chemical composition of minority phase could be taken as an indicator for the presence of the two phases.

\begin{figure}
\begin{center}
\includegraphics[width=\columnwidth]{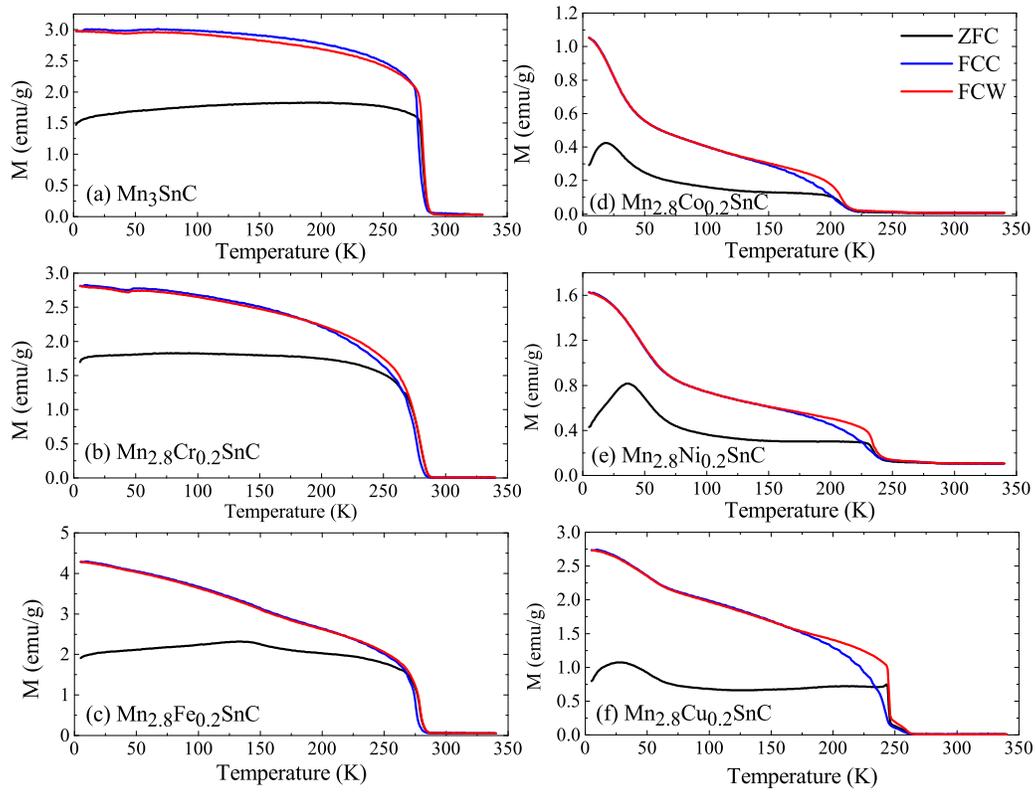}
\caption{Magnetization as a function of temperature of (a) Mn$_{3}$SnC, (b) Mn$_{2.8}$Cr$_{0.2}$SnC, (c) Mn$_{2.8}$Fe$_{0.2}$SnC, (d) Mn$_{2.8}$Co$_{0.2}$SnC, (e) Mn$_{2.8}$Ni$_{0.2}$SnC and (f) Mn$_{2.8}$Cu$_{0.2}$SnC polycrystalline samples.}
\label{fig:4TrMMT}
\end{center}
\end{figure}

The observations so far raise two fundamental questions, (a) what is the cause for the formation of impurity phases in Co, Ni and Cu doped Mn$_3$SnC? and (b) is there any interaction between the two majority and minority phases in these multiphasic compositions? To answer the second question first, magnetization as a function of temperature recorded on all five doped samples and compared with that of Mn$_3$SnC. The ZFC, FCC and FCW curves of magnetization are depicted in figure \ref{fig:4TrMMT}(a) to (f). Magnetization curves of the Cr and Fe doped samples are similar to that of undoped Mn$_3$SnC (figure \ref{fig:4TrMMT}(a)). The first order magnetic transformation from paramagnetic state to a state with complex magnetic order occurs at almost similar temperature in all the three compounds, and these are listed in Table \ref{tab1}. The transformation temperature is also in agreement with reported work on Fe substitution in Mn$_3$SnC \cite{wang2010322m}. The two noticeable changes in the magnetization curves are the transition hysteresis which is quite large in the case of Cr doped sample (figure \ref{fig:4TrMMT}(b)) and a hump-like feature at around 130K in the ZFC curve of Fe doped sample as can be seen in figure \ref{fig:4TrMMT}(c).

In contrast, the substitution of Co, Ni and Cu at Mn site of Mn$_{3}$SnC produces drastic alterations in the transformation temperature. Although all these three are multiphasic compounds with antiperovskite phase being the majority phase, the first order transformation temperature reduces drastically in comparison with the undoped compound as can be seen from figures \ref{fig:4TrMMT}(d) to (f) for Mn$_{2.8}$Co$_{0.2}$SnC, Mn$_{2.8}$Ni$_{0.2}$SnC and Mn$_{2.8}$Cu$_{0.2}$SnC respectively. The surprising fact is that none of the magnetic phases identified from XRD have their ordering temperature around the observed transition temperature. Such a shift in the transition temperature indicates presence of magnetic interaction between the majority and minority magnetic phases.  Additionally, below 50 K, a broad hump-like feature is evident in the ZFC curve of all these three compounds.  This feature disappears in the field cooled magnetization curves hinting towards a possible non-ergodic transition in these three compounds. Mn$_3$Sn$_{0.5}$Ga$_{0.5}$C which has a cluster glassy ground state also exhibits a similar feature in its ZFC magnetization curve which disappears in the field cooled curves \cite{dias201795ph}. Another interesting observation is the scaling of transformation temperature with the atomic number of the transition metal atom. As Z changes from 27 (Co) to 29 (Cu), the transformation temperature increases from 220K to 264K.

\begin{figure}
\begin{center}
\includegraphics[width=\columnwidth]{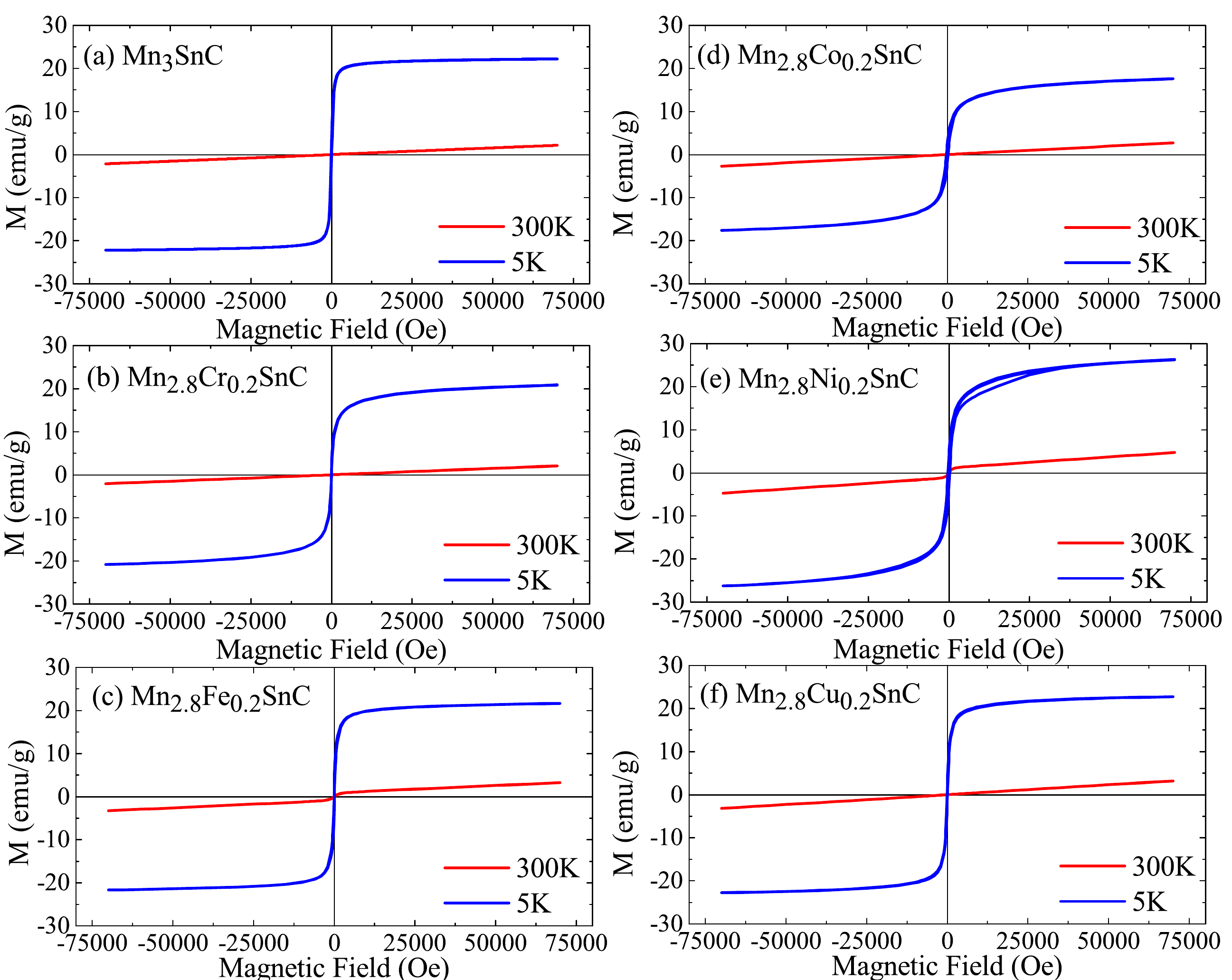}
\caption{Isothermal magnetization at 5K and 300K recorded in a magnetic field of $\pm$ 7T for (a) Mn$_{3}$SnC, (b) Mn$_{2.8}$Cr$_{0.2}$SnC, (c) Mn$_{2.8}$Fe$_{0.2}$SnC, (d) Mn$_{2.8}$Co$_{0.2}$SnC, (e) Mn$_{2.8}$Ni$_{0.2}$SnC and (f) Mn$_{2.8}$Cu$_{0.2}$SnC polycrystalline samples.}
\label{fig:5TMMH}
\end{center}
\end{figure}

To further check the nature of magnetism in these compounds isothermal magnetization data recorded on all the compounds at 300 K and 5 K and are presented in figure \ref{fig:5TMMH}. The samples were cooled in zero field to the recording temperature and the field was ramped to $\pm$ 7 T. At 300K, all compounds, except two, Mn$_{2.8}$Fe$_{0.2}$SnC and Mn$_{2.8}$Ni$_{0.2}$SnC exhibit paramagnetic behavior. In Fe doped and Ni-doped compounds, a minor ferromagnetic character is visible. The observation of ferromagnetic nature in Mn$_{2.8}$Ni$_{0.2}$SnC can be understood from the fact that the Heusler phase, Mn$_2$NiSn has a T$_C$ of around 530 K \cite{lakshmi200225hy}. But the Fe doped compound is supposed to be structurally single-phase and hence observation of ferromagnetic character in isothermal magnetization at 300 K is surprising. At 5K, all compounds exhibit hysteresis loops typical of ferrimagnetic compounds. Though the initial rise of magnetization appears to be slower in Cr and Fe doped compounds than that in Mn$_3$SnC, the magnetization values at 7 T are nearly similar for all the three compounds. The slower growth of magnetization and ferromagnetic signal at 300K in Fe doped compound points towards presence of competing magnetic interactions. In the case of multiphasic compounds, the magnetization at 7T attains its lowest value in Mn$_{2.8}$Co$_{0.2}$SnC and then with the change in transition metal atom from Co to Ni to Cu, it increases systematically in a way very similar to the transformation temperature.

\begin{figure}
\begin{center}
\includegraphics[width=\columnwidth]{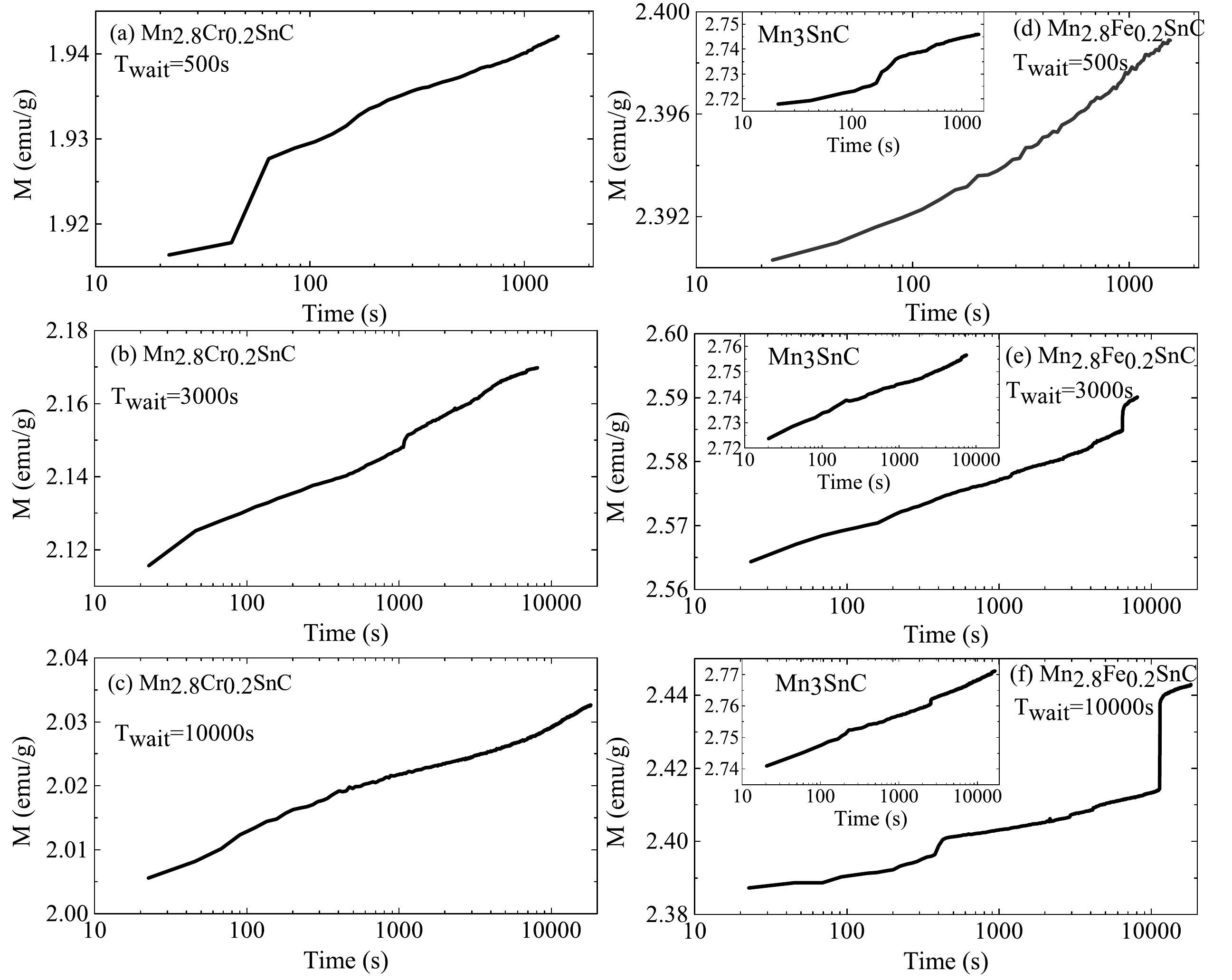}
\caption{(a)--(f) Magnetization as a function of time of Mn$_{2.8}$Cr$_{0.2}$SnC and Mn$_{2.8}$Fe$_{0.2}$SnC, polycrystalline samples recorded at 5K after wait time t$_{w}$ of 500s, 3000s, 10000s. Inset of figure (d), (e), (f) is the time-dependent magnetization of Mn$_{3}$SnC observed at corresponding wait time t$_{w}$.}
\label{fig:8TMMtime1}
\end{center}
\end{figure}

\begin{figure}
\begin{center}
\includegraphics[width=\columnwidth]{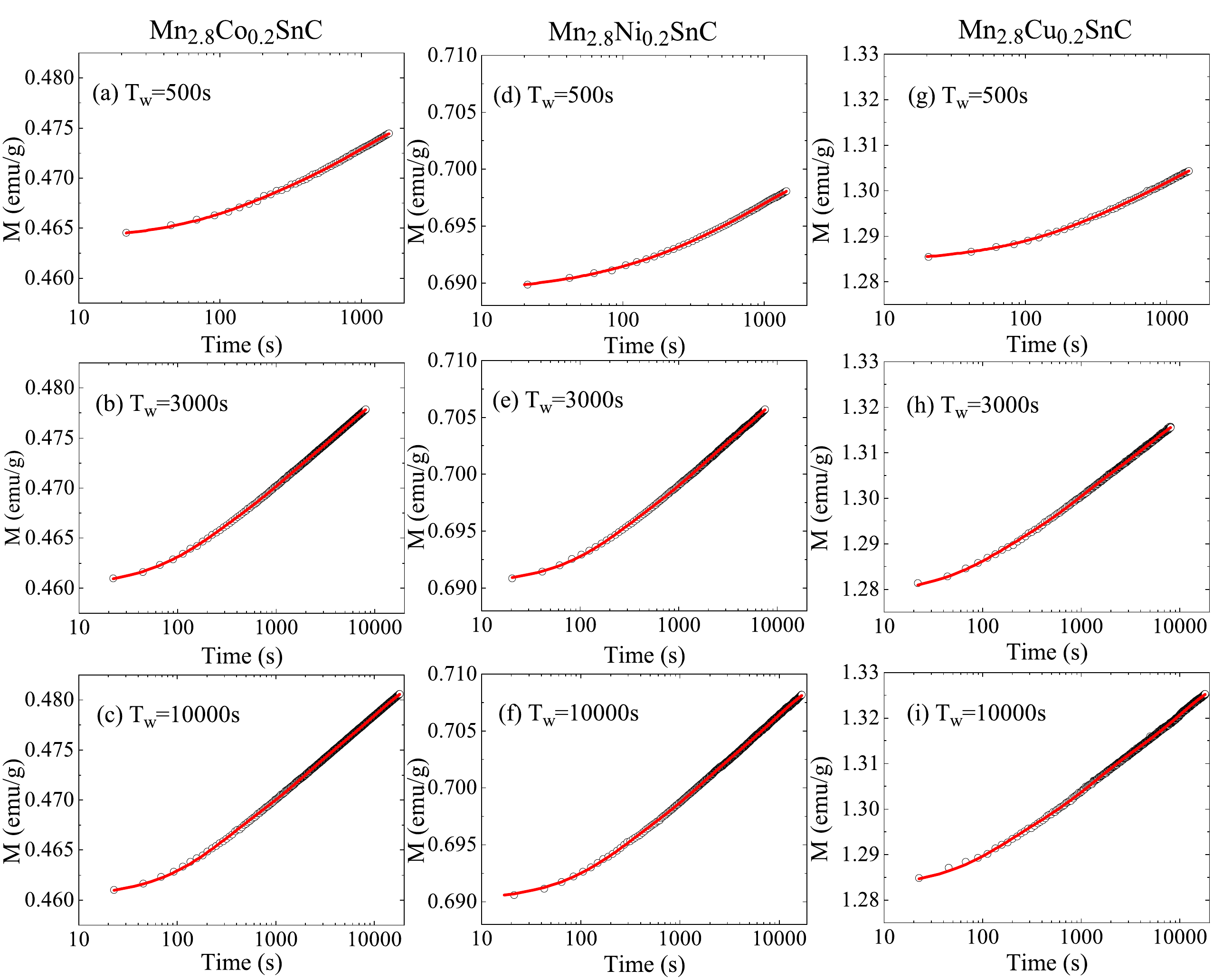}
\caption{(a)--(i) Time dependent magnetization plots recorded at 5K for Mn$_{2.8}$Co$_{0.2}$SnC, Mn$_{2.8}$Ni$_{0.2}$SnC and Mn$_{2.8}$Cu$_{0.2}$SnC after 3 different wait times of 500s, 3000s and 10000s.  Solid lines are fits to a third order exponential growth function implying presence of more then one relaxation time.}
\label{fig:9TMMtime2}
\end{center}
\end{figure}

To understand the origin of magnetic interactions in single phase and multiphase, transition metal-doped Mn$_3$SnC, magnetization was recorded as a function of time. For these measurements, the samples were cooled in zero field down to 5K and after a fixed wait time (T$_{wait}$) of 500s, 3000s or 10000s, a field of 100 Oe was applied and magnetization was recorded as a function of time. Such magnetization curves for all the two single-phase samples along with pristine Mn$_3$SnC are presented in figure \ref{fig:8TMMtime1} and for the three multiphase compounds in figure \ref{fig:9TMMtime2}. In the case of Mn$_3$SnC magnetization follows the expected exponential growth for all three wait times as can be seen from the curves presented as insets in figure \ref{fig:8TMMtime1}. In addition to this exponential increase, small magnetization steps are visible especially for shorter T$_{wait}$ = 500s. With addition of Cr into the antiperovskite lattice, the magnetization step becomes more prominent but overall character of exponential growth of magnetization is preserved. The only other noticeable aspect is the change in magnetization value with wait time. In Mn$_3$SnC, the variation of magnetization irrespective of the wait time is similar. However, in Mn$_{2.8}$Cr$_{0.2}$SnC, the magnetization increases from about 1.915 emu/g for T$_{wait}$ = 500s (figure \ref{fig:8TMMtime1}(a)) to 2.11 emu/g for T$_{wait}$ = 3000s (figure \ref{fig:8TMMtime1}(b)) at the start of observation time. When T$_{wait}$ is increases further to 10000s the magnetization decreases slightly to 2.05 emu/g (figure \ref{fig:8TMMtime1}(c)). A similar variation of magnetization values is seen in the case of Mn$_{2.8}$Fe$_{0.2}$SnC (figure \ref{fig:8TMMtime1}(d)--(f)) with magnetization initially increasing from 2.39 emu/g for T$_{wait}$ = 500s to 2.56 emu/g for T$_{wait}$ = 3000s and then decreasing back to 2.39 emu/g for T$_{wait}$ = 10000s. Additionally a prominent magnetization step is seen at an observation time of 10000s in case of T$_{wait}$ = 10000s. At nearly same observation time, a similar step but of much smaller magnitude is also seen for T$_{wait}$ = 3000s.

Magnetization steps have a different origin. The fact that such steps are not seen in isothermal magnetization curves of respective compounds and that the compounds do not have any other magnetic impurity phase at least to the level seen from x-ray diffraction, the presence of steps in magnetization versus time data depict orientation of all magnetic domains in favorable direction. The magnetic properties of Mn-based antiperovskite carbides are susceptible to carbon stoichiometry \cite{Dias2014363ef,gaonkar2019mo}. Such regions, if present, could have their spins pinned in a different direction to start with which rotate in the favorable direction after lapse of a particular time. In the case of Mn$_{2.8}$Fe$_{0.2}$SnC, presence of magnetic impurity which pins the spins of some domains in a different orientation cannot be ruled out. A broad hump at around 130K seen in ZFC temperature-dependent magnetization curve and a weak hysteresis loop seen in isothermal magnetization data at 300K supports the presence of magnetic impurity in Fe doped Mn$_3$SnC. Presence of such magnetic impurity could also be the reason for observed variation of magnetization value with wait time.

For the three multiphase compounds, Mn$_{2.8}$Co$_{0.2}$SnC, Mn$_{2.8}$Ni$_{0.2}$SnC and Mn$_{2.8}$Cu$_{0.2}$SnC, the time evolution of magnetization is completely different from the above three compounds and are presented in figure \ref{fig:9TMMtime2}(a)--(c) respectively. Here the magnetization of all three compounds at all wait times does not display a linear behavior on a logarithmic time scale indicating the presence of more than one relaxation time scales. The observed magnetization behavior can be fitted with an exponential growth function of order three. The second difference is that there are no magnetization steps or a variation in the magnitude of magnetization as a function of wait time. The starting value of magnetization at all wait times is nearly similar for any of the three compounds. However, here too as observed for isothermal magnetization, the starting value of magnetization increases with the increase in Z number of the doped transition metal atom.

The results so far indicate two different types of compounds - crystallographically single-phase and multiphase. In single-phase compounds the observed magnetic properties are similar to undoped Mn$_3$SnC. Though, subtle differences, especially in time-dependent magnetization exist and point to existence of magnetic phase separation. In multiphase compounds, the magnetic properties exhibit drastic changes compared to Mn$_3$SnC. Such a change in magnetic properties is surprising because the doping level of the transition metal impurity is only about 7\%. These vast changes in magnetic properties indicate strong magnetic interactions between the major antiperovskite phase and minor impurity phase or phases. Further, most of the magnetic properties seem to scale with Z number of the doped transition metal atom. Therefore, to understand the nature of magnetic interactions between different magnetic phases, frequency-dependent ac susceptibility measurements have been carried out on all multiphase compounds. These measurements have also been extended to single-phase compounds including undoped Mn$_3$SnC.

\begin{figure}
\begin{center}
\includegraphics[width=\columnwidth]{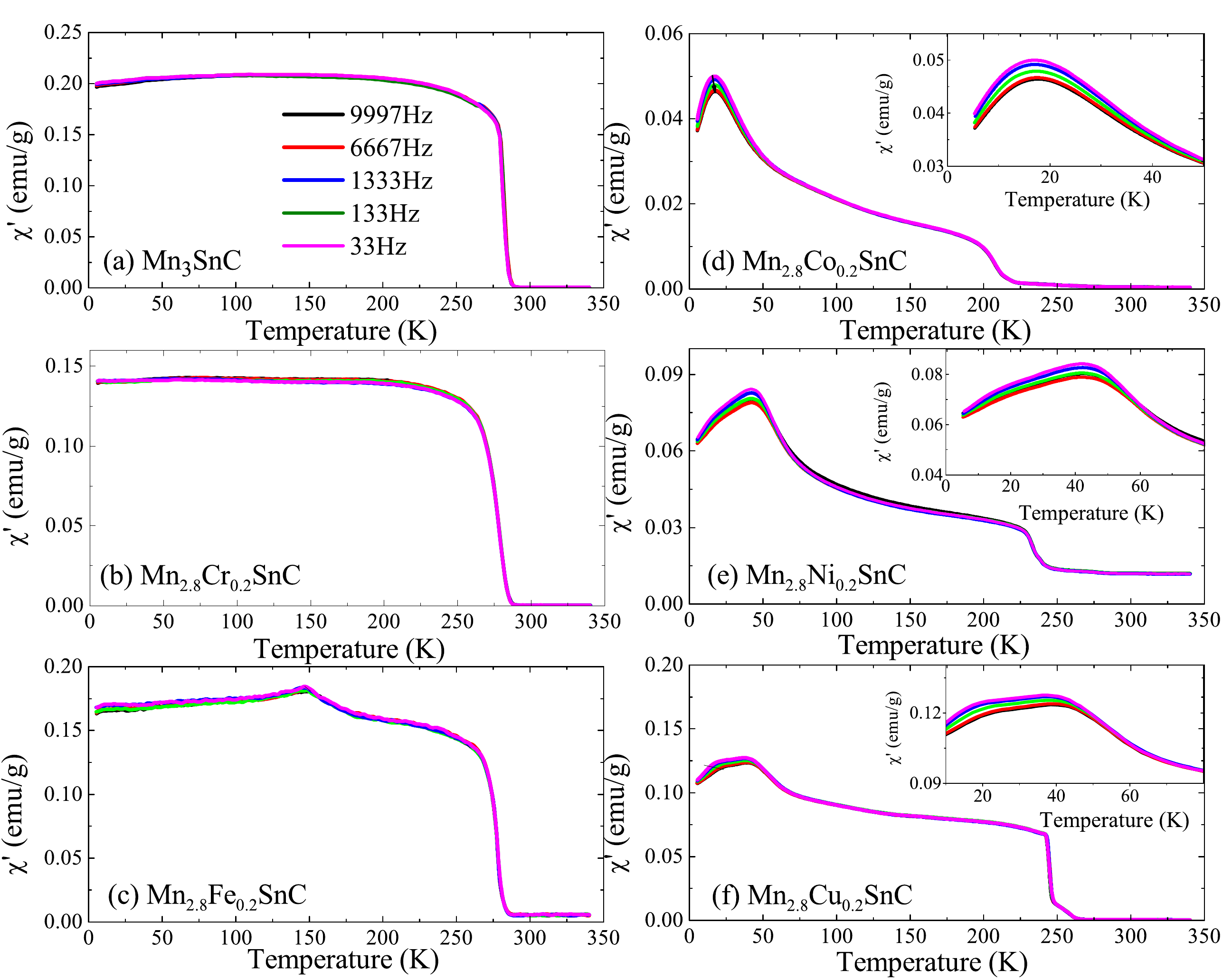}
\caption{AC susceptibility (real part) $\chi'$ measured for (a) Mn$_{3}$SnC, (b) Mn$_{2.8}$Cr$_{0.2}$SnC, (c) Mn$_{2.8}$Fe$_{0.2}$SnC, (d) Mn$_{2.8}$Co$_{0.2}$SnC, (e) Mn$_{2.8}$Ni$_{0.2}$SnC and (f) Mn$_{2.8}$Cu$_{0.2}$SnC samples at different frequencies.}
\label{fig:6TMfreqdep}
\end{center}
\end{figure}

\begin{figure}
\begin{center}
\includegraphics[width=\columnwidth]{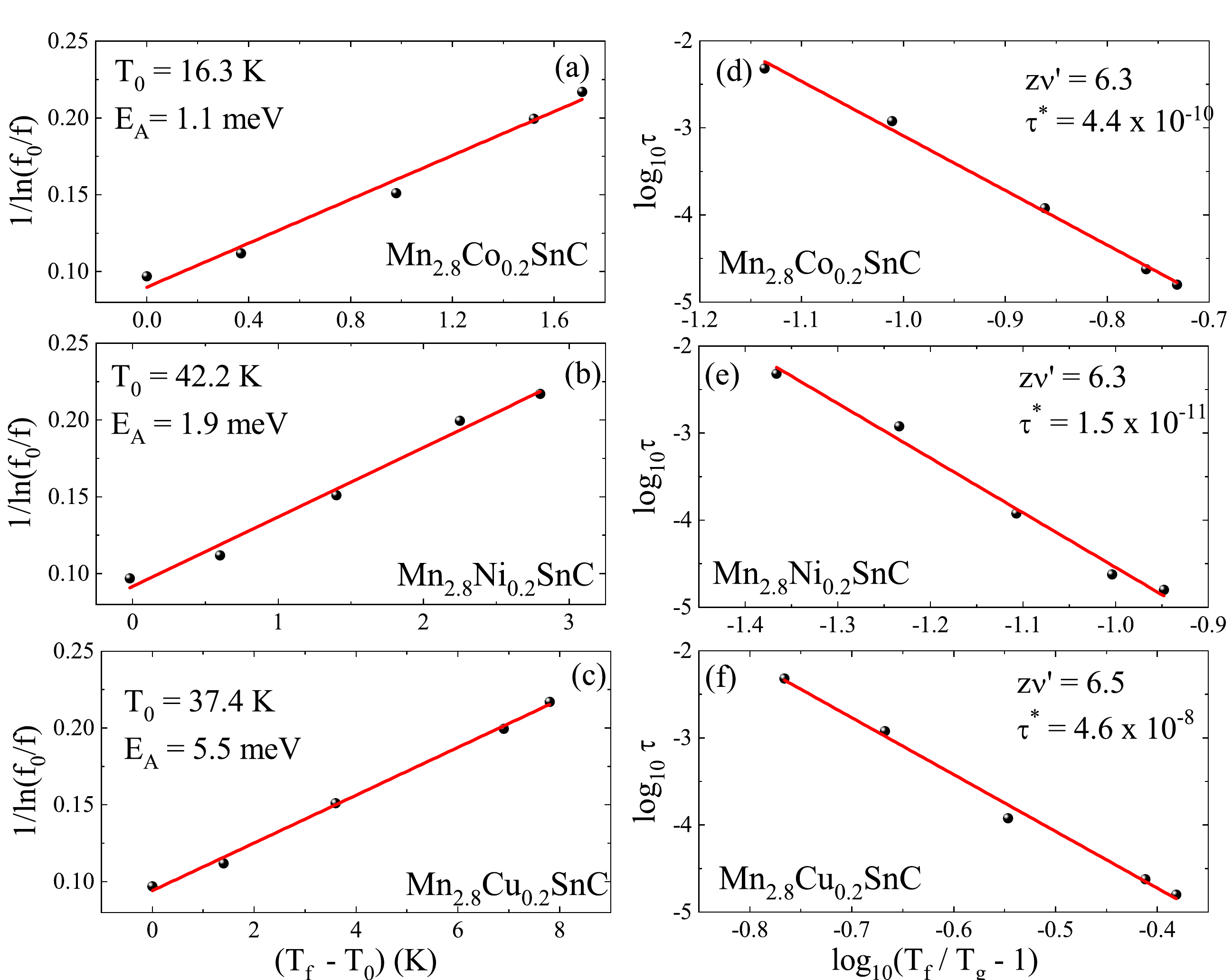}
\caption{Frequency dependence of temperature in Mn$_{2.8}$Co$_{0.2}$SnC, Mn$_{2.8}$Ni$_{0.2}$SnC and Mn$_{2.8}$Cu$_{0.2}$SnC samples obtained from Ac susceptibility data fitted using Vogel-Fulcher empirical law and the characteristic Critical slowing down Avramov-Milchev equation.}
\label{fig:7VFL-CSD}
\end{center}
\end{figure}

The real part of ac susceptibility signal ($\chi'$) recorded for all six compounds as a function of temperature at five different frequencies between 33 Hz and 10000 Hz is plotted in figure \ref{fig:6TMfreqdep}. In the three single-phase compounds, Mn$_3$SnC, Mn$_{2.8}$Cr$_{0.2}$SnC and Mn$_{2.8}$Fe$_{0.2}$SnC, $\chi'$ do not any dependence of frequency while for the multiphase compounds, Mn$_{2.8}$Co$_{0.2}$SnC, Mn$_{2.8}$Ni$_{0.2}$SnC and Mn$_{2.8}$Cu$_{0.2}$SnC a clear dependence on frequency is observed around the low-temperature maxima seen in ZFC magnetization data of these compounds. The appearance of such broad maxima in ZFC curve at low temperature along with a deviation between FC and ZFC curves are considered as typical signatures of a glassy ground state \cite{dias201795ph,wu20065ma}. Though a similar maximum, but a much higher temperature, is seen in Mn$_{2.8}$Fe$_{0.2}$SnC, no frequency dependence of $\chi'$ is observed. To confirm the presence or absence of a glassy state in each of the three compounds, the position of low-temperature maxima was extracted and plotted as a function of applied frequency. It can be seen that the position of maxima, T$_f$ systematically decreases in amplitude and shifts towards higher temperature with increasing applied frequency. This variation of T$_f$ as a function of frequency can be well described by Vogel-Fulcher relation as can be seen from figure \ref{fig:7VFL-CSD} and thus confirming the presence of a frozen magnetic glassy state. The estimated values of relative shift in spin freezing temperature, $\delta T_{f} = \Delta T_{f}/T_{f} \Delta \ln\nu$ in Mn$_{2.8}$Co$_{0.2}$SnC, Mn$_{2.8}$Ni$_{0.2}$SnC and Mn$_{2.8}$Cu$_{0.2}$SnC are estimated and listed in Table \ref{tab2}. These values are much smaller than the reported value for a typical superparamagnetic system, $\alpha$-(Ho$_{2}$O$_{2}$(B$_{2}$O$_{3}$)) ($\delta$T$_{f} \sim 0.28$)  and slightly larger than that reported for canonical spin-glass system like CuMn ($\delta T_{f} = 0.005$) \cite{mydosh1993sp}. These observations suggest that the glassy state in these Co, Ni and Cu doped Mn$_3$SnC is not of atomic origin rather arises from a cluster of atoms \cite{chakrabarty2014cl}. Hence the ground state in these three compounds can be identified as a cluster glass.

In figure \ref{fig:6TMfreqdep}(d), (e) and (f) Avramov-Milchev expression for critical slowing down of the relaxation time, $\frac{\tau}{\tau_0}$ = $({\frac{T_{f}}{T_{0}}-1})^{-z\nu^{'}}$ is fitted to the experimental data of Co, Ni and Cu doped Mn$_3$SnC respectively. Here, $\tau$ represents dynamical fluctuation time scale corresponding to the observation time $\frac{1}{2\pi\nu}$, $\tau_0$ is the microscopic relaxation time, T$_{f}$ is the freezing temperature at specific observation time while T$_{0}$ is the transition temperature equivalent to T$_{f}$ as $\nu$ $\rightarrow$ 0 and $z\nu$ is the dynamical exponent \cite{chakrabarty2014cl}. The values of $\tau_0$ and ${z\nu^{'}}$ obtained from the fitting the data of Co, Ni and Cu doped Mn$_3$SnC are listed in Table \ref{tab2}. The values obtained for all the three compounds indicate presence of strongly interacting spin clusters.

\begin{table*}
\centering
\caption{\label{tab2} Dynamical exponent $z\nu'$ and Microscopic relaxation time $\tau_0$ obtained from critical slowing down analysis of frequency dependent ac susceptibility data recorded at five different frequencies between 33 Hz to 10000 Hz for Mn$_{2.8}$Co$_{0.2}$SnC, Mn$_{2.8}$Ni$_{0.2}$SnC and Mn$_{2.8}$Cu$_{0.2}$SnC samples.}
\begin{center}
\begin{tabular*}{\textwidth}{@{}lccccc}
\hline
Samples & \multicolumn{2}{c}{Vogel-Fulcher parameters} & Relative shift & \multicolumn{2}{c}{Avramov-Milchev parameters} \\
 & Spin-Glass & Activation & in spin freezing & Dynamical & Microscopic \\
 & temperature & energy  & temperature & exponent & relaxation time \\
 & T$_{0} (K)$ & E$_{A} (meV) $ & $\delta T_{f} (K)$ & $z\nu'$ & $\tau_{0}$ \\
\hline
Mn$_{2.8}$Co$_{0.2}$SnC & 16.3 & 1.1 & 0.0178 & 6.3 & 4.4 $\times$ 10$^{-10}$ \\
Mn$_{2.8}$Ni$_{0.2}$SnC & 42.2 & 1.9 & 0.0117 & 6.3 & 1.5 $\times$ 10$^{-11}$ \\
Mn$_{2.8}$Cu$_{0.2}$SnC & 37.4 & 5.5 & 0.0365 & 6.5 & 4.6 $\times$ 10$^{-8}$ \\
\hline
\end{tabular*}
\end{center}
\end{table*}

The structural and magnetic studies on transition metal-doped Mn$_3$SnC show that the higher Z transition elements do not readily form an antiperovskite phase and instead phase separate into multiphase compounds comprising of transition metal-rich Heusler phase and a transition metal deficient antiperovskite phase. The phase separation seems to be related to the filling up of the $3d$ band. Less than half-filled $d$ band elements like Cr are easily incorporated into the antiperovskite lattice of Mn$_3$SnC as a substituent for Mn atom. While more than half-filled $d$ band elements like Co, Ni and Cu do not substitute Mn in Mn$_3$SnC resulting in multiphase compounds consisting among others an Heusler phase. Although structural phase separation is not clearly seen in Fe doped compounds, its magnetic properties point towards magnetically phase-separated ground state. The observation of weak ferromagnetic hysteresis at 300K, well above the transition temperature of the antiperovskite phase of Mn$_{2.8}$Fe$_{0.2}$SnC, a broad peak like feature in the ZFC magnetization data, magnetization steps are all indicators of possible magnetic phase separation.

Interestingly, in the multiphase compounds there is a significant magnetic interaction between the majority and minority phases. The transition temperature of the antiperovskite phase is reduced from 280 K in Mn$_3$SnC to about 210 K in Mn$_{2.8}$Co$_{0.2}$SnC. Though it increases to about 245 K in Mn$_{2.8}$Cu$_{0.2}$SnC, there are indicators of multiple magnetic transitions, probably of the minority phases in the magnetization data. Most prominent among these is the occurrence of a broad maximum at low temperatures (T $<$ 50K), which is seen to exhibit frequency dependence per Vogel-Fulcher law and a cluster glassy ground state.

\section{Conclusion}
In conclusion, the transition metal-doped antiperovskites Mn$_{2.8}$T$_{0.2}$SnC (T = Cr, Fe, Co, Ni and Cu) exhibit interesting structural and magnetic properties. While Cr and Fe doped compounds are a structurally single phase with magnetostructural transformation temperature very close to that of undoped Mn$_3$SnC, the Co, Ni and Cu doped compounds exhibit structural phase separation and a drastic decrease in the transformation temperature. Detailed investigation of structural and magnetic properties of these transition metal-doped compounds reveal that with increase in atomic number of the transition element, there is a tendency to push carbon out of the lattice resulting in formation of Heusler type and DO19 type impurity phases. Interestingly, an interaction between majority antiperovskite phase and the minor impurity phases ensues leading to observation of non-ergodic behavior of magnetization and a cluster glassy ground state in all of the phase-separated compounds.

\section*{Acknowledgement}
VNG thanks University Grants Commission for BSR Fellowship. Financial support from Science and Engineering Research Board (SERB), DST under the project EMR/2017/001437 is gratefully acknowledged.

\bibliographystyle{elsarticle-num}
\bibliography{refer}
\end{document}